\documentclass{aa}  

\usepackage{graphicx}
\usepackage{txfonts}
\begin{document}

   \title{Structure formation with primordial black holes to alleviate early star formation tension revealed by JWST}


   \author{P. E. Colazo.\fnmsep\thanks{E-mail: patricio-c@hotmail.com}
          \inst{1,2}
          \and
F. Stasyszyn
          \inst{2,3}
          \and
N. Padilla
          \inst{2,3}
           }

   \institute{Facultad de Matemática, Astronomía, Física y Computación, UNC, Argentina\\
         \and
             Instituto de Astronomía Teórica y Experimental, CONICET--UNC, Argentina\\
         \and
             Observatorio Astronómico de Córdoba, UNC, Argentina\\
             }
   \date{Received 21 March 2024; accepted 09 April 2024}

  \abstract
{This Letter explores the potential role of primordial black holes (PBHs) to address cosmological tensions as the presence of more massive than expected galaxies at high redshifts, as indicated by recent James Webb Space Telescope observations.} 
{Motivated by inflation models that enhance the power at scales beyond the observable range that produce PBHs with Schechter-like mass functions, we aim to explain the excess of high redshift galaxies via a modification of the $\Lambda$ cold dark matter power spectrum that consists in adding (i) a blue spectral index $n_b$ at $k_{\text{piv}}=10/$Mpc and (ii) Poisson and isocurvature contributions from massive PBHs that only make up $0.5\%$ of the dark matter.}
{We simulated these models using the SWIFT code and find an increased abundance of high redshift galaxies in simulations that include PBHs. We compared these models to estimates from  James Webb Space Telescope observations.}
{Unlike the $\Lambda$ cold dark matter model, the inclusion of PBHs allowed us to reproduce the the observations with reasonable values for the star formation efficiency.
Furthermore, the power spectra we adopted potentially produce PBHs that can serve as seeds for supermassive black holes with masses $7.57 \times 10^4 M_{\odot}$.}
{}
\titlerunning{Primordial black holes to alleviate the early star formation tension }

   \maketitle
%

\section{Introduction}

Primordial black holes (PBHs) have garnered significant attention due to their potential candidacy as a solution to the enigma of dark matter (DM). Several studies have investigated the properties and challenges of PBHs within the framework of the standard cosmological model of cold dark matter (CDM) including a cosmological constant, $\Lambda$CDM \citep{Carr_2023, Sureda_2021}.

The quest to identify the components of the DM has focused on various candidates, including axions, weakly interacting massive particles (WIMPs), fermionic particles, or neutralinos \citep{Olive_2014}. Many of these candidates require new physics beyond the known standard model. However, PBHs have emerged as a natural solution that could have been produced during the early universe due to the effect of cosmic inflation on primordial inhomogeneities \citep{Inomata_2017, Clesse_2015}.

Even though PBHs were thought to have low abundances after searches for compact objects in the Milky Way \citep{Alcock_machos}, the detection of gravitational wave signals from black hole mergers by the LIGO-Virgo-KAGRA collaboration \citep{abbott_2023} with black hole masses in slight conflict with those expected from stellar remnants reignited interest in PBHs \citep{Carr_1974, Zel'dovich, Hawking_1971, Carr_2023, boyuan_2022_pbh_halos}.

Consequently, in the past few years, authors have imposed constraints on the abundance of PBHs using different observables. We can categorise the possible PBH mass functions into two main types. The first is the monochromatic case, which is often used for  simplicity. The second is the extended mass function, which offers a broader distribution of masses and has gained increasing attention recently \citep{Carr_2023,Sureda_2021}.

The community has imposed various constraints on the fraction of PBHs in DM, employing different mass distribution models. For example, \cite{Padilla_2021} constrained both monochromatic and extended mass distributions with Poisson effects on the large-scale structure, while other authors applied multiple constraints using different techniques \citep{Auclair_2023, NANOGrav_2021, Carr_2017, Sasaki_2018, Niikura_2019}. Recent updates on the extended and monochromatic cases can be found in \cite{Carr_2021} and \cite{Sureda_2021}.

The inclusion of PBHs in cosmology has far-reaching implications. One notable application is the explanation of gravitational wave signals detected by LIGO through PBH mergers \citep{Bird_2016, Sasaki_2016, Raidal_2017}. They have also been proposed as a solution to the baryon asymmetry problem \citep{Ambrosone_2022}, and they could serve as seeds for the origin of magnetic fields in the universe \citep{Araya_2021,Papanikolaou2023PhRvD}.  They may also have implications for resolving the Hubble tension and explaining gravitational waves in the nano-Hz range \citep{Li_2023}. Other notable application postulates PBHs as an explanation for the stochastic gravitational wave background \citep{Agazie_2023,Yi_2023}.
Primordial black holes could also have the potential to address the core-cusp controversy \citep{Boldrini_2020, boyuan_2022_stars, Kashlinsky_2021}. 

Primordial black holes can also impact the formation of cosmic structures, accelerating the collapse compared to the $\Lambda$CDM model \citep{Carr_2020,Inman_2019,boyuan_2022_stars}. Particularly, PBHs could contribute to the current paradigm by explaining the presence of quasars at $z > 7$. The presence of PBHs has implications for the early formation of galaxies and may help resolve recent controversies related to observations by the James Webb Space Telescope (JWST) \citep{Gouttenoire_2023, Su_2023, boyuan_2022_pbh_halos,Goulding2023,Kokorev2023}. 

The first images obtained from JWST through the Cosmic Evolution Early Release Science (CEERS) survey pose a potentially significant challenge to the $\Lambda$CDM cosmological model. \cite{Labbe_2023} identified candidate massive galaxies with stellar masses exceeding $10^{10} M_{\odot}$ at redshifts $ z \geq 7.4$, with one candidate potentially reaching a stellar mass of $\sim 10^{11} M_{\odot}$.  
These selections were updated with spectroscopic confirmation, which alleviate but still do not resolve the tension with $\Lambda$CDM  \citep{Parashari_2023}.

Other authors have identified similar high redshift galaxy candidates \citep{Yan_2023,Harikane_2022,Naidu_2022}, and recent studies have employed these estimates to test scenarios involving a blue-tilted primordial power spectrum as possible alternatives that produce higher abundances of massive galaxies at high redshifts \citep{Parashari_2023,Brummel_2023,boyuan_2022_pbh_halos}. 

This Letter shares a similar perspective while introducing a different, small modification to the power spectrum beyond the observable range consisting of the presence of PBHs at below the percent level, considering the Poisson effect from massive PBHs along with isocurvature effects (see \citealt{boyuan_2022_pbh_halos}) and adopting Press-Schechter mass functions for PBHs.

This article is structured as follows. In  Section \ref{sec:simulations}, we describe the details of our simulations and numerical methods. Section \ref{sec:results} presents our results for the abundance of high redshift galaxies, while Section \ref{sec:discussion} provides an approximate calculation that highlights the potential significance of primordial black holes as seeds of supermassive black holes (SMBHs) within our Universe. We conclude in Section \ref{sec:conclusions}.

\begin{table}
    \centering
    \renewcommand{\arraystretch}{1} 
    \begin{tabular}{@{}ccc@{}}
        \hline
        Parameter & Description &value\\
        \hline
        $N_{\text{DM}}$ & Number of
particles of DM& $1024^3$ \\
        $L_{\text{box}}$ & Box side length& $205\ \rm{cMpc / h}$ \\
        $\text{soft}_{\text{com}}$ & Softening& $14.8\ \text{kpc}$ \\
        $z_{\text{init}}$ & Initial redshift & $127$\\
        $\Omega_{\text{cdm}}$ & Dark matter density parameter& $0.267$ \\
        $\Omega_{\text{b}}$ & Baryon density parameter& $0.049$  \\
        $\Omega_{\Lambda}$ & Dark energy density parameter& $0.684$ \\
        $\sigma_{8}$ & RMS of Power spectrum at 8 Mpc/h& $0.8118$  \\
        $h$ & Hubble parameter& $0.673$  \\
\hline
\ &Extra parameters in the PBH sim.& \\
\hline
        $k_{\text{piv}}$ & Pivot scale & $10\ \rm{cMpc}^{-1}$ \\
        $n_b$ & Blue spectral index & $2.5$ \\
        $f_{\text{PBH}}$ & Fraction of PBH in
dark matter& $0.005$\\
        $M^*_{\text{PBH}}$ & Characteristic PBH
        mass& $10^4\ M_{\odot}$ \\
    \end{tabular}
    \caption{Simulation and cosmological parameter values.}
    \label{tab:parametros_simu}
\end{table}

\section{Simulations}
\label{sec:simulations}

In this work we aim to compare the abundance of galaxies detected by JWST with that expected in simulations. The first simulation used is a $\Lambda$CDM model with cosmological parameters based on Planck2018 \citep{Planck_2020}, as our fiducial model. The second simulation includes PBH effects in the initial power spectrum.
We assume PBHs make up $0.5\%$ of the DM in this simulation and that they follow a Press-Schechter mass function with a characteristic mass of $10^4M_\odot$, which formed in the fixed conformal time (FCT) case \citep{Sureda_2021} with a blue index $n_b=2.5$.
In both cases the simulations were performed using the SWIFT simulation code  \citep{swift}.
The parameters of the simulations are outlined in Table \ref{tab:parametros_simu}. 

We aim to compare the number density of  JWST-detected galaxies at high redshifts with those expected in $\Lambda$CDM models and when adding the slight modification of including a sub-percent mass in PBHs.

To generate the initial conditions, we used a power spectrum that - in the case of the PBH model - incorporates modifications that (i) would give rise to the formation of PBHs in the early universe and (ii) modifications due to the PBHs themselves. 
 We used the second-order Lagrangian perturbation theory (2LPT) to convert the initial power spectra into  initial particle distributions.

\textit{Initial and evolved power spectra}. Our PBH simulation follows a power spectrum corresponding to the $\Lambda$CDM one up to $k_{\text{piv}}$, a scale deliberately chosen to lie beyond the observable range.  We chose $k_{\text{piv}} \approx 10 \rm{Mpc}^{-1}$ as in \cite{Sureda_2021}. The primordial modifications we applied make, at most, a $15\%$ change in amplitude at this particular scale. 

This choice of pivot scale is the one that produces the strongest acceptable change in the power spectrum. The choice of blue index $n_{\text{b}}$ influences the tilt in the exponential drop-off of the PBH mass function; however, its effect is only marginal on the power spectrum. Conversely, $f_{\text{PBH}}$ holds greater sway over the results as it affects both the amplitude and the break in the mass distribution. However, we opted for the maximum value allowed by current constraints for an FCT mass function. The horizon crossing mass function, alternative to FCT, is steeper for the same blue index, producing a smaller effect than our chosen mass function.

The mass distribution of PBHs that form in the early universe depends on the primordial power spectrum as well as on the timing when the amplitude of the linear fluctuation associated with PBH formation is assessed.

We use the following primordial power spectrum: 
$$P_{\mathrm{primordial}}(k) = A_S {\left( \frac{k}{k_0} \right)}^{n_s}, $$
where $A_s$ denotes the characteristic amplitude of fluctuations at $k_0 = 0.05 \rm{\rm{Mpc}}^{-1}$, and $n_s$ represents the spectral index, often referred to as the red index, due to its proximity to and slight deviation from $1$. 

We introduce a new spectral index denoted as $n_b$:

\[
\Tilde{P}_{\mathrm{primordial}}(k) =
\begin{cases}
    A_S \left( \frac{k}{k_0} \right)^{n_s}, & \text{si } k < k_{\mathrm{piv}}  \\
    A_S \varepsilon \left( \frac{k}{k_0} \right)^{n_b}, & \text{si } k > k_{\mathrm{piv}} ,
\end{cases}
\]
\label{equ:broken_power}
where $\varepsilon$ serves as a normalisation factor. 
When adopting a blue index $n_b=2.5$, the chosen power spectrum is able to produce PBHs characterised by a Press-Schechter mass function with characteristic mass $M_* = 10^4 M_\odot$. \cite{Sureda_2021} used different observables to constrain the abundance of PBHs with extended mass functions, and find that in our adopted case they can make up $0.5 \times 10^{-2}$ of the DM at most. We set the fraction of DM in PBHs, $f_{\text{PBH}}$, to this value.

The presence of PBHs, characterised by their discrete and massive nature, introduces a significant Poisson effect on the gravitational potential capable of modifying the evolution of cosmic fluctuations \citep{Padilla_2021}, which we added to the initial power spectrum.

We additionally incorporated the effect of isocurvature perturbations due to the presence of massive PBHs on the universe. We  followed \citep{boyuan_2022_pbh_halos}, where the growth factor for these perturbations is
\begin{equation*}
    \begin{aligned}
D(a) & \simeq\left(1+\frac{3 \gamma}{2 a_{-}} s\right)^{a_{-}}-1, \quad s=\frac{a}{a_{\mathrm{eq}}} \\
\gamma & =\frac{\Omega_{\mathrm{m}}-\Omega_{\mathrm{b}}}{\Omega_{\mathrm{m}}}, \quad a_{-}=\frac{1}{4}(\sqrt{1+24 \gamma}-1),
\end{aligned}
\end{equation*}
where the scale factor at matter radiation equality is $a_{eq} = 1 / 3401$. The final isocurvature power spectrum follows 
\begin{equation*}
\begin{aligned}
P_{iso}(k) & \simeq\left[ \bar{f} D(a)\right]^{2}/\Bar{n}_{PBH},
\end{aligned}
\end{equation*}
where $\bar{f}$ represents the fraction of mass density in black holes with the same number density as JWST galaxies.

The complete power spectrum, including these considerations, reads as follows:
$$P(k,z) = \Tilde{P}_{\mathrm{primordial}}(k)\ T^2(k)\ D_1^2(z) + f_{PBH}^2 P_{Poisson}^{PBH}(k,z) + P_{iso} (k,z),$$
where $T(k)$ is the transfer function and $D_1$ corresponds to the growth factor.

\begin{figure}
    \centering
    \includegraphics[width=0.45\textwidth]{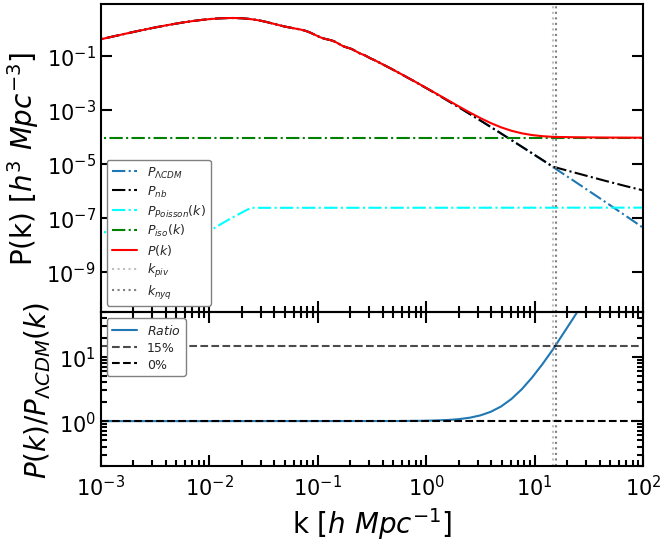}\vskip-.3cm
    \caption{Power spectra adopted in the construction of initial conditions (top).
    The blue line shows the initial spectrum for the $\Lambda CDM$ model.  The initial power spectrum with all contributions from PBHs is shown in red.   The black line shows the broken power spectrum, the cyan line shows the contribution due to the the Poisson effect, and the green line shows the isocurvature effect. The vertical dash-dotted grey lines show the location of $k_{\text{piv}}$ and $k_{\text{nyq}}$. Bottom: Ratio between the power spectra with and without PBH effects.
    }
    \label{fig:Power_spectrum_construction}
\end{figure}

Figure \ref{fig:Power_spectrum_construction} presents an illustration of the initial power spectrum incorporating the presence of PBHs. The initial $\Lambda$CDM power spectrum without PBHs at $z = 127$ is displayed in blue.
The red line shows the final power spectrum for PBHs with all contributions (primordial power spectrum, the contribution of blue index in black, the Poisson effect in cyan, and the isocurvature effect in green). The lower panel shows the ratio between the power spectrum for the case including PBHs (red) and 
the power spectrum without PBH effects to highlight the contribution of the Poisson  noise and isocurvature of 
$\sim 13\%$ at $k_{piv}$ and $\sim 15\%$ at $k_{nyq}$.

We used these power spectra to construct the initial conditions  at $z = 127$, for which we used the N-Genic code \citep{Springel_2015}. The most significant properties of these simulations are presented in Table \ref{tab:parametros_simu}. We used the Rockstar code to identify haloes in our simulations \citep{Behroozi_2012}.

Using the SWIFT code, we ran both simulations down to a final redshift of $z=9$.

\section{Results}
\label{sec:results}
We measured the abundance of high mass galaxies from the $\Lambda$CDM simulation and from the simulation that includes PBHs.  
To this end, we used the cumulative mass function for halos and employed a simple conversion to estimate the total stellar mass contained within them
as follows \citep{boyuan_2022_pbh_halos}:
\begin{equation*}
\begin{aligned}
    M_*=\epsilon \Omega_b/\Omega_m M_{\rm halo}, 
\end{aligned}
\end{equation*}
where $\epsilon$ is the efficiency of star formation, $\Omega_b$ is the baryon density parameter, and $\Omega_m$ is the matter density parameter (See Table \ref{tab:parametros_simu}). Reasonable values for the efficiency of star formation for $\Lambda$CDM models lie around $\epsilon=0.2$, with possible extreme values of up to $\epsilon=0.32$ (\citealt{Gribel}, see also \citealt{Behroozi, Tacchella}).

\begin{figure}
    \centering
    \includegraphics[width=0.43\textwidth]{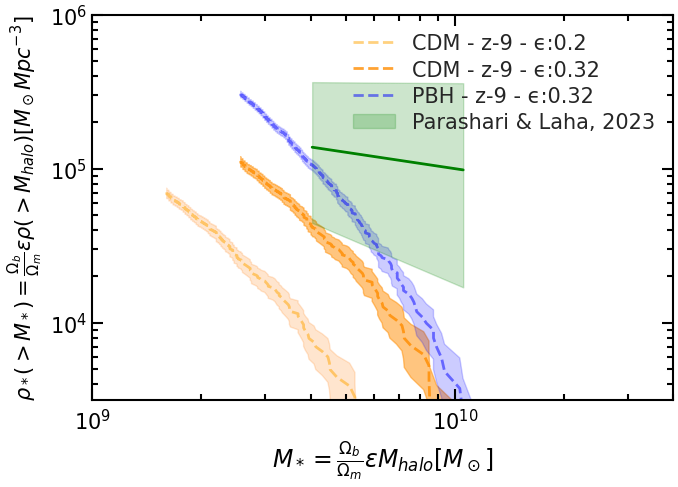}\vskip-.3cm
    \caption{
    Cumulative stellar mass function at $z=9$ for JWST galaxies \cite{Parashari_2023} based on photometric redshift estimates updated with spectroscopy (green-filled area).  The $\Lambda$CDM case is shown in orange and the model with a small contribution from primordial black holes is in blue; both have an efficiency of star formation  $\epsilon=0.32$. Shaded regions show the Poisson error for each simulation. Light orange illustrates the CDM case with a more typical value of efficiency $\epsilon=0.2$.}
    \label{fig:funcion_masa}
\end{figure}

Figure \ref{fig:funcion_masa} shows the cumulative density of haloes exceeding a specified stellar mass at redshift $z=9$ for different values of star formation efficiency. The orange and blue lines correspond to the CDM and PBH simulations, respectively, and the shaded regions show Poisson errors. The green-shaded region shows the JWST galaxy abundance estimate from \cite{Parashari_2023} based on photometric redshift estimates updated with spectroscopy.

In the context of the CDM model, a moderate star formation efficiency ($\epsilon=0.2$) fails to explain the abundances detected by JWST, while the upper limit of $\epsilon\leq0.32$ 
allows only a marginal capacity of the model to reproduce high redshift galaxy abundances.

The introduction of PBHs provides a compelling solution. The early presence of PBHs could influence feedback mechanisms and could accelerate and enhance the efficiency of star formation \citep{boyuan_2022_stars}. This, in principle, could allow higher efficiency values, rendering the adopted $\epsilon=0.32$ more realistic.

It is essential to emphasise that the behaviour of PBH models depends on a combination of parameters, namely $n_b$, $k_{\text{piv}}$, and $f_{\text{pbh}}$. Only the latter was chosen so that the associated abundance of PBHs with this particular mass function is the highest possible within current observables. 

Our results illustrate how implementing a relatively minor adjustment to the fiducial model can help alleviate a particular tension.  We subsequently explore whether this particular choice of modelling offers further solutions to current open questions.  

\section{Discussion}
\label{sec:discussion}

In this work we have attempted to show how PBHs can contribute to solve the JWST galaxy abundance within a cosmological context \citep{Carr_2023}, without trying to pinpoint the best parameters of the PBH model adopted. While we have taken prior research into consideration, such as the PBH population parameters guided by \citep{Sureda_2021}, it is important to emphasise that our focus lies on the use of the proposed cosmology rather than its accuracy. 

To put this into context, a similar change in the abundance of high redshift galaxies to that obtained by increasing the efficiency by a rather significant $50\%$ can be obtained by simply increasing $\sigma_8$ by a mere $2\%$ over the Planck 2018 value.  This would allow $\Lambda$CDM to  produce a similar abundance as the PBH model presented here. However, this would alter cluster abundances at $z=0$; we would replace one tension with a different one. Therefore, this approach seems less suitable. In contrast, the introduction of PBHs accelerates structure growth in early times allowing the JWST data to be fit with $\epsilon=0.32$ without affecting the late time cluster abundance (or $\sigma_8$).

Primordial black holes could help to address other problems within the $\Lambda$CDM paradigm.  Even though most galaxies are thought to harbour central SMBHs our current understanding struggles to provide mechanisms that give rise to these  objects. While we have a grasp of how black holes with masses ranging from $8$ to $100\,M_{\odot}$  form from massive stars, the formation of SMBHs remains a puzzle. Furthermore, recent measurements of high redshift quasars continue to add complexity to these ideas. 

For instance, ’J1007+2115’, identified as the most massive known quasar with a black hole mass of $M_{BH} \approx 1.5 \times 10^9\,M_{\odot}$ at redshift $z > 7.5$ (\citealt{Feng_2021}, \citealt{Yang_2020}), presents a conundrum. Estimates suggest a seed black hole mass ($M_{\text{seed}}$) of approximately $10^4\,M_{\odot}$ at redshift $z \sim 30$ is needed. However, there is uncertainty surrounding this estimate because alternative models demand even higher seed masses, of the order of $10^{5} - 10^{6}\,M_{\odot}$, based on the direct collapse of baryonic gas.

Another example is presented by \citet{Goulding2023}, who show that the required high redshift black hole to stellar mass ratio should be higher by a factor of two to three than in the local Universe, implying a possibly higher influence of the seed mass on the evolution of high redshift black holes.

Other studies, such as  \cite{Johnson_2013}, have delved into radiation feedback within the first generation of stars and how these affect accretion processes with implications regarding seed black hole mass predictions. They concluded that the required seed mass to explain observations should be $M \gtrsim 10^5\,M_{\odot}$, a prediction that aligns with the wider range of $10^4-10^6\,M_{\odot}$ predicted by direct collapse scenarios.

The extended mass function of PBHs allows the possibility of providing different origins for SMBHs. For instance, as shown by \cite{Yifan}, the evaporation of PBHs in the early universe could  help prevent molecular hydrogen from forming which would allow gas clouds to undergo direct collapse and form SMBHs.

Our choice of PBH model presents another alternative for the origin of SMBHs, as it not only helps explain the abundance of high-z galaxies, it may also help to produce seeds for SMBHs. As mentioned earlier, various authors have suggested the necessity of seed black holes with masses  $M_{BH} \sim 10^4 M_{\odot}$ to account for the presence of SMBHs in the high redshift Universe  \citep{Johnson_2013,Feng_2021,Yang_2020,Kokorev2023}.

We estimate the minimum mass of PBHs that could act as seeds for SMBHs  by simply demanding they have at least the same number density of present-day galaxies with $> 10^{10}$ solar masses.  This choice is justified since these are all thought to host SMBHs in their centres. It is worth noting that massive PBHs do not experience significant growth through accretion in their early stages. Therefore, their mass distribution is practically the same as  at their formation.
At later stages when these objects fall into the central regions of halos, they are indeed expected to show significant mass growth via mergers with other central black holes of accreted galaxies. 

We integrated our PBH mass function and determined the black hole mass such that it yields a similar number density to that of galaxies with $10^{10} M_{\odot}$ of $\simeq 7 \times 10^{-4} h^3 $\rm{Mpc}$^{-3}$ \citep{Ross_2015}, 
$$ n(>M_{\text{pbh}}) = \int_{M_{\text{min}}}^\infty \frac{dn}{dM} \, dM\,$$
where $dn/dM$ is the differential mass function of primordial black holes obtained via the Press-Schechter formalism. We obtain $ M_{min} = 7.57 \times 10^4 M_{\odot}$, which is stable under PBH mass function parameters for similar $M_{\rm PBH}^*$. This  matches previous estimates for the SMBH seed mass surprisingly well \citep{Johnson_2013,Feng_2021}, which place seeds within the range of mass $ [ 10^4-10^6 M_{\odot}]$. 

One could consider the possibility that one of these potential PBH seeds, given their random nature that gives rise to their Poisson contribution to the potential, initially resided at valleys in the density distribution, away from peaks that will collapse into galaxies.  PBHs form as they come into causal contact shortly after the Big Bang; they are initially high density peaks, but the evolution of fluctuations within the horizon can essentially reshuffle the locations of peaks that later form galaxies.  However, even if they are located away from late time peaks, they can still source isocurvature fluctuations that affect matter and baryons and can produce additional collapsed objects. At late times this can favour merging  with other black holes of stellar or primordial origin, furthering the growth of its associated overdensity. One would be tempted to expect galaxies to form around these PBHs; however, their total contribution to the galaxy power spectrum is quite small.  Consequently, it is safe to expect that massive PBHs lie mostly near density peaks that give rise to galaxies, and that they fall to their centres due to physical mechanisms such as dynamical friction and interactions with stars and gas.  The few extragalactic massive PBHs will likely be accreted by galaxies, and eventually merge with their central PBHs.

It would be interesting to detect SMBHs outside the nuclei of galaxies in their process of infall towards the new host galaxy, but this effort is still ongoing \citep{vanDokkum_2023, Tremmel_2018}. Even though their detection would  not constitute proof that PBHs are the seeds of SMBHs, it would indeed help confirm SMBH growth via mergers with other central BHs.

Even though the abundance of PBHs in our chosen model fits the expected abundance of SMBH seeds quite well, there could be other phenomena producing seeds such as the mechanism proposed by \citep{Yifan}. Still, having PBHs that expedite the formation of early halos may also facilitate the earlier merger of stellar black holes, potentially lowering the mass of the required seeds.

\section{Conclusions}
\label{sec:conclusions}

We have studied the potential role of PBHs constituting a small sub-percent fraction of the DM in the abundance of high redshift massive galaxies as observed by the JWST. Our results show that a small abundance of PBHs can serve as a viable addition to conventional DM models to explain the presence of these high redshift objects, without modifying our underlying understanding of cosmological structure formation. Our findings neither exclude other possible astrophysical effects, such as the evolution of star formation physics between the early and late Universe, nor the intrinsic uncertainties involved in the interpretation of JWST observations. 

We used the SWIFT code to run two simulations, a $\Lambda$CDM model and an additional one with effects of PBHs making up a small, sub-percent contribution to the DM, included in the initial power spectrum.

We find that the PBH model  alleviates, to some degree, the challenge posed by the presence of high redshift galaxies.  Using a  star formation efficiency  of $\epsilon=0.32$, which is the extreme value found in CDM studies \citep{Gribel}, we are able to reach similar levels of galaxy abundances as found with JWST observations by \cite{Parashari_2023}.  This high star formation efficiency could be more easily attainable when including PBHs due to the early collapse they induce in their surroundings \citep{boyuan_2022_stars}.

We also explored the potential of PBHs as seeds for the formation of SMBHs. For the PBH mass function adopted here, PBHs with $M > 7.57 \times 10^4 M_{\odot}$ reproduce the observed number density of galaxies with masses exceeding $10^{10} M_{\odot}$.  This seed mass is within the required levels to explain the abundance of high redshift SMBHs \citep{Goulding2023,Kokorev2023}.

If the need for the small fraction of DM in black holes adopted here was confirmed, this could serve as a way to constrain primordial power spectrum parameters such as $n_b$ and $k_{\text{piv}}$ which, in turn, would help constrain the parameter space of inflation models.

\begin{acknowledgements}
PEC acknowledges support from a PhD fellowship from CONICET-Argentina.  NDP was supported by a RAICES, a RAICES-Federal.  NDP, PC and FAS received support PICT-2021-I-A-00700 from the Ministerio de Ciencia, Tecnología e Innovación, Argentina.   FAS thanks support by grants PIP 11220130100365CO,
PICT-2016-4174, PICT-2021-GRF-00719 and Consolidar-2018-2020, from CONICET, FONCyT (Argentina) and SECyT-UNC.
This work used computational resources from CCAD-UNC, which is part of SNCAD-MinCyT, Argentina. Thanks to Dr. Nicola Menci for reviewing the manuscript.
\end{acknowledgements}
%
%

\bibliographystyle{aa} 
\bibliography{output}

\end{document}